# Crowding induces entropically-driven changes to DNA dynamics that depend on crowder structure and ionic conditions

**Warren M. Mardoum[1], Stephanie M. Gorczyca [1], Kathryn E. Regan[1], Tsai-Chin Wu[1], and Rae M. Robertson-Anderson*[1]**

[1]Department of Physics and Biophysics, University of San Diego, San Diego, CA, 92110, USA

**\*Correspondence:** randerson@sandiego.edu

## Abstract

Macromolecular crowding plays a principal role in a wide range of biological processes including gene expression, chromosomal compaction, and viral infection. However, the impact that crowding has on the dynamics of nucleic acids remains a topic of debate. To address this problem, we use single-molecule fluorescence microscopy and custom particle-tracking algorithms to investigate the impact of varying macromolecular crowding conditions on the transport and conformational dynamics of large DNA molecules. Specifically, we measure the mean-squared center-of-mass displacements, as well as the conformational size, shape, and fluctuations, of individual 115 kbp DNA molecules diffusing through various *in vitro* solutions of crowding polymers. We determine the role of crowder structure and concentration, as well as ionic conditions, on the diffusion and configurational dynamics of DNA. We find that branched, compact crowders (10 kDa PEG, 420 kDa Ficoll) drive DNA to compact, whereas linear, flexible crowders (10 kDa, 500 kDa dextran) cause DNA to elongate. Interestingly, the extent to which DNA mobility is reduced by increasing crowder concentrations appears largely insensitive to crowder structure (branched vs linear), despite the highly different configurations DNA assumes in each case. We also characterize the role of ionic conditions on crowding-induced DNA dynamics. We show that both DNA diffusion and conformational size exhibit an emergent non-monotonic dependence on salt concentration that is not seen in the absence of crowders.

## 1    Introduction

Biological cells are highly crowded by macromolecules of varying sizes and structures. This complex crowded environment has been shown to directly impact key DNA processes and functions including replication, transcription, transformation, gene expression, and chromosomal compaction [1-6]. Investigating the impact of crowding on DNA is further motivated by the design of gene therapy and drug delivery systems, as well as the production and manipulation of synthetic cells and nanomaterials [2, 3, 7, 8]. Crowding can induce changes in DNA conformations, such as compaction, swelling, and elongation; and alter its diffusivity and intramolecular fluctuations [9-11]. However, the exact effect that crowding has on DNA mobility and conformation remains poorly understood. The wide range of differing results presented in the literature likely stems from the myriad of sizes and types of crowders, as well as the varying ionic conditions, used in *in vitro* experiments – both of which directly impact the effect of crowding on DNA.

Because crowding studies are largely motivated by the role crowding plays in cells, several studies have investigated DNA dynamics *in vivo* [1, 5, 12-14]. While these studies can directly illuminate DNA behavior in cells, the role that each variable (e.g., crowder size, structure, concentration, ionic condition) plays in the measured dynamics is hard to discern. Thus, researchers have turned to *in vitro* studies to methodically explore the role of each variable separately [1, 15, 16]. Most *in vitro* crowding studies use synthetic polymers, similar in size to the majority of small proteins in cells, at concentrations similar to those found in cells (20 – 40% w/v). Polysaccharides, such as dextran, polyethylene glycol (PEG) and Ficoll, are advantageous as crowders because they are inert, nonbinding, commercially available in a range of molecular weights, and can often be described by basic polymer theory [17-23]. While these crowders are often used interchangeably to mimic cellular crowding, they differ considerably in their structure and conformational shape.

Dextran is a linear, flexible polymer which is reported to assume a random coil conformation in solution with an empirical scaling of hydrodynamic radius with molecular weight of $R_h \sim M_w^{0.49}$ [20, 24, 25]. While this scaling exponent is close to that of an ideal polymer chain (scaling exponent of 0.5), dextran coils have been shown to be highly asymmetric [20]. At high concentrations, dextran can form entanglements more easily than branched crowders of similar molecular weight due to its linear, flexible structure. Further, its asymmetric shape suggests that at high enough concentrations nematic ordering is also more readily accessible than for branched crowders [26]. PEG is a flexible linear polymer with a hydrodynamic radius that scales as $R_h \sim M_w^{0.56}$. However, at concentrations above ~7% w/v it self-associates to form highly branched structures which are suggested to behave more like hard spheres than random coils [27]. Ficoll is a highly-branched polymer that assumes a compact spherical conformation that can be described by the empirical scaling $R_h \sim M_w^{0.43}$ [19, 21].

A number of previous studies have examined the effects of these crowders on polymer transport, conformation, and stability [28-37]. However, the role these crowders play in the diffusion or conformation of large DNA molecules remains largely unexplored. Further, no previous studies have directly compared the impact that each of these distinctly-shaped crowders has on DNA dynamics. One previous study investigated the role of crowder shape on protein diffusion by measuring the diffusion of heart-shaped BSA and Y-shaped IgG proteins in crowded solutions of either BSA or IgG [38]. Results showed that the varying excluded volume resulting from the differently-shaped proteins played the most important role in diffusion, as opposed to crowder concentration or direct interactions. A previous simulation study examined the effect of crowder shape on the conformation of stiff rod polymers [26]. Results showed that at high concentrations, spherical crowders caused compaction of the polymers, reducing their radius of gyration $R_g$, while spherocylindrical particles increased $R_g$. The increase in $R_g$ was thought to be caused by nematic ordering of spherocylindrical particles which allow the polymers to elongate in the direction of ordering.

We previously measured the diffusion and conformational dynamics of DNA crowded by dextran of varying molecular weights and concentrations [39, 40]. We found that the decrease in DNA diffusivity with increasing dextran concentration was actually less than expected based on the increasing viscosity of the crowding solutions [39]. Namely, measured diffusion coefficients followed a weaker scaling with viscosity than the classical Stokes-Einstein scaling $D \sim \eta^{-1}$. This "enhanced" diffusion was coupled with conformational elongation of DNA from its dilute random coil configuration.

While entropically-driven depletion interactions tend to compact large macromolecules to maximize the available volume for the small crowders, elongation or swelling could ensue if energetic or enthalpic effects counteract depletion forces. Because DNA is negatively charged, entropy maximization competes with the electrostatic cost of compaction, which can lead DNA to preferentially elongate rather than compact to reduce its conformational volume in certain cases. However, varying the ionic conditions of the solution can directly tune the electrostatic cost of compaction via positive ions (e.g., $Na^+$) screening DNA charge. In fact, increased salt concentrations have been shown to be essential to PEG-induced condensation/aggregation of DNA (known as $\Psi$- compaction) [34, 35, 41, 42]. Since the discovery of Ψ-compaction [41], there have been numerous experimental [4, 6, 8, 42-52] and theoretical [53-59] studies investigating this phenomenon. Experiments have shown that Ψ-compaction of 166 kbp DNA is a first-order transition from random coil configurations to compacted states; and when the concentration and/or molecular weight of PEG is increased the required NaCl concentration for compaction is reduced [42]. The lowest reported NaCl concentration to induce compaction was 50 mM, which required 25% w/v of 11.5 kDa PEG, in contrast to the 300 mM NaCl required when 10% w/v of 3 kDa PEG was used. Magnetic and optical tweezers studies have examined the force required to uncoil compacted 48.5 kbp DNA in the presence of 15-30% w/v PEG and varying concentrations of NaCl [35, 44]. These studies found that this force increased as the PEG molecular weight and NaCl concentration increased (up to 6 kDa PEG and 2 M NaCl). The remaining studies on DNA compaction have focused on the effects of salt in confined conditions [6, 43, 60-64] or in the presence of charged crowders [65-72]. These previous studies have shed important new light on the role that charge screening plays in crowding-induced DNA compaction. However, the interplay between crowder shape and salt concentration has been left completely unexplored. Further, these studies have all focused on steady-state DNA conformations, so the question of how ionic strength impacts the mobility and conformational fluctuations of crowded DNA remains an important open question.

Here, we explore the effect of crowder shape and ionic conditions on the intriguing crowding-induced DNA dynamics that have previously been reported. Specifically, we track the center-of-mass mean-squared displacements, as well as the size and shape of single DNA molecules diffusing in solutions of crowders with varying structural properties and molecular weights (10 kDa dextran, 500 kDa dextran; 10 kDa PEG, 420 kDa Ficoll). Based on the literature, throughout the text we classify PEG and Ficoll as branched crowders and dextrans as linear crowders. We further explore the effect of salt concentration on the measured DNA properties by examining cases of 5-fold increased and decreased NaCl under maximally crowded conditions. We show that crowding-induced DNA conformations are highly-dependent on crowder structure, with branched crowders compacting DNA while linear crowders induce elongation. We further show DNA diffusion exhibits a surprising non-monotonic dependence on salt concentration in crowded environments.

## 2    Methods and Materials

All experimental methods and computational analysis, briefly summarized below, are thoroughly described and verified in [39, 40, 73]. Linear 115 kbp DNA molecules are prepared through replication of supercoiled bacterial artificial chromosomes in Escherichia coli, followed by extraction, purification and enzymatic linearization [40]. A trace amount of DNA is fluorescent-labeled with YOYO-I (Thermo Scientific) and embedded in a solution of 0-40% w/v crowding polymers dissolved in aqueous buffer (10 mM Tris-HCl, 1 mM EDTA, 10 mM NaCl, 4% β-mercaptoethanol). The four different crowders used are: 10 kDa dextran (Sdex), 500 kDa dextran (Ldex), 420 kDa Ficoll, and 10 kDa PEG (all purchased from Sigma Aldrich). Unless indicated, NaCl concentration is 10 mM. The viscosity of each crowded solution was measured using optical tweezers microrheology as described previously [40, 74].

To determine the diffusion coefficients as well as the conformational size, shape and fluctuations of crowded DNA, single embedded DNA molecules were imaged for 30 seconds at 10 frames per second using a high-speed CCD camera on a Nikon A1R epifluorescence microscope with 60x objective. All presented data are for ensembles of >200 molecules. Using custom-written algorithms (Matlab), we track the center-of-mass (COM) position $(x, y)$, as well as the lengths of the major and minor axes ($R_{max}$ and $R_{min}$, respectively) of each molecule over time. We calculate the COM mean-squared-displacement in the $x$ and $y$ directions ($<\Delta x^2>$, $<\Delta y^2>$) to determine the diffusion coefficient $D$ via $<\Delta x^2> = <\Delta y^2> = 2Dt$. Error bars are calculated using bootstrapping for 1000 sub-ensembles [75]. We use $R_{max}$ as a measure of conformational size and we quantify the degree of molecular elongation or asymmetry by comparing $R_{max}$ to $R_{min}$. Finally, we determine the time-dependence and lengthscales of intramolecular state fluctuations by calculating the ensemble-averaged fluctuation range $f(t) = </R_{max}(0)-R_{max}(t)/>/<R_{max}>$ for a given lag time $t$. $f(t)$ can be understood as the relative lengthscale (compared to the conformational size) over which a given molecule fluctuates or "breathes" between different conformational states during a time $t$.

## 3    Results and Discussion

### 3.1    Crowder Structure

We first examine the role of crowder structure on DNA diffusion and conformation. Figure 1 displays the dependence of measured DNA diffusion coefficients on crowder concentration and solution viscosity for the four different crowders. As shown, the diffusion coefficients, $D$, for all crowders decrease with increasing crowder concentration, as expected given the increasing viscosity, $\eta$, of crowded solutions. However, the mobility reduction does not scale with increasing viscosity as classically predicted by the Stokes-Einstein relation, $D=k_bT/6\pi\eta R$. Instead, the scaling of $D$ with $\eta$ is shallower than $\eta^{-1}$ with an average scaling for all crowders of $D\sim\eta^{-0.5}$. Only for the two highest Ldex concentrations, does this enhanced diffusion effect begin to be suppressed, with diffusion values beginning to more closely align with Stokes-Einstein predictions. The viscosity at these concentrations are also much higher than those measured for the other crowders at the same concentrations. This effect is likely due to the linear dextran polymers forming entanglements. In contrast, branched crowders of similar molecular weight assume hard sphere conformations, preventing entanglements. Sdex also does not form entanglements because of its smaller size compared to Ldex. For reference, the hydrodynamic radius, $R_h$, of Sdex, Ldex, PEG and Ficoll are 2.2 nm, 15.5 nm, 3.2 nm, and 5.6 nm, respectively [17, 19, 20, 25]. The conformational size of Ldex in solution is >3$x$ larger than the other crowders despite being similar molecular weight to Ficoll, thereby promoting polymer overlap and entanglement. Another interesting result that can be seen in Figure 1B is the similarity between the viscosities of PEG and Ficoll at equal concentrations, as well as their effect on DNA diffusion, despite their large $M_w$ difference and ~2$x$ difference in $R_h$. This result suggests that PEG is indeed self-associating and forming branched structures which more closely mimic the branched structure and size of Ficoll. We note that the enhanced diffusion effect is more pronounced for linear crowders (SDex, Ldex) than branched crowders (PEG, Ficoll), which likely arises from differences in the crowding-induced DNA conformations which we describe below.

We previously attributed enhanced diffusion to conformational changes, which reduce the effective volume of the DNA [39, 40]. To build on this assumption and determine the role that crowder structure plays in conformational changes, we look to the measured major and minor axis lengths $R_{max}$ and $R_{min}$. Figure 2(A-D) shows the distributions of DNA major axis lengths for each crowder type and concentration. There is a noticeable difference between the distributions for branched crowders compared to linear crowders. While branched crowders reduce $R_{max}$ (narrowing distributions and shifting them to the left), linear crowders tend to increase $R_{max}$ (widening and shifting distributions to the right).

To quantify the degree to which the average major axis length $<R_{max}>$ as well as the spread in the distributions (quantified by the standard deviation $\Delta R_{max}$) vary with crowding, we plot $<R_{max}>$ vs $\Delta R_{max}$ (Fig 2E). The difference between linear and branched crowders is evident: branched crowders decrease $\Delta R_{max}$ and $<R_{max}>$, signifying compaction, while linear crowders tend to increase both quantities, indicating elongation or swelling. Further, elongated or swollen configurations access a wider range of states (larger $\Delta R_{max}$) than dilute condition random coil states, while the range of accessed states for compacted configurations is reduced. This reduction is indicative of ordered compaction, in which tight intramolecular packing causes the molecules to fluctuate between fewer states. We note that for Sdex $R_{max}$ actually decreases slightly, but $\Delta R_{max}$ still increases appreciably and the distributions show a more pronounced large $R_{max}$ tail compared to dilute conditions. Thus, we interpret this data as still demonstrating elongation or swelling rather than compaction. Finally, the ($<R_{max}>$, $\Delta R_{max}$) data points for all crowders show no discernible trend with crowder concentration, with most data points for 10-40% w/v clustering together. These data indicate that the crowding-induced conformational changes are an all or nothing effect, similar to the discrete first-order phase transition observed in $\Psi$-compaction.

To delineate between symmetric swelling versus elongation (induced by linear crowders), and determine if compacted configurations (induced by branched crowders) are more ellipsoid or spherical in nature than dilute condition random coils, we compare $R_{max}$ to $R_{min}$ (Fig 2F). The larger the $R_{max}$:$R_{min}$ ratio, the more asymmetric the conformation is. As shown, dilute condition random coil configurations exhibit a ~3:2 aspect ratio, as previously predicted and shown [76, 77]. Conversely, compacted conformations induced by branched crowders are more spherical in nature, similar to hard spheres. Linear crowders, on the other hand, markedly increase $R_{max}$:$R_{min}$, signifying elongation rather than symmetric swelling.

Because the spread in our conformational state distributions ($\Delta R_{max}$) vary for the different crowders, we investigated whether this spread is a result of a heterogeneous ensemble of DNA molecules each assuming a different static conformation, or whether the ensemble is fairly uniform but each molecule transitions between different states over time. To determine the extent to which single molecules transition or "breathe" between different conformational states over time, we measure the change in $R_{max}$ for varying lag times $t$ and normalize by $<R_{max}>$. We refer to this quantity as the fluctuation range, $f(t)=</R_{max}(0)-R_{max}(t)/>/<R_{max}>$. As shown in Figure 3(A,B), $f(t)$ increases over time for all cases and approaches a steady-state value, which we term the steady-state fluctuation range $f_{ss}$ (Fig 3C). Thus the spread in the conformational state data (Fig 2) arises from single molecules breathing between different states over time, rather than a heterogeneous ensemble of static conformations. The results shown in Figure 3 also display a marked difference between linear and branched crowders: while Sdex and Ldex increase $f_{ss}$ by ~30% from the dilute value, both PEG and Ficoll reduce $f_{ss}$ to ~50% of the dilute value. These data validate our interpretation that elongated molecules access a broader range of conformational states compared to more rigid compact configurations. It is also notable that while there is a stark difference between the effect of linear crowders versus branched crowders, the fluctuation length shows little dependence on the concentration or size of crowders. This result is similar to the discrete phase transition seen in Figure 2, and underlines the critical role that crowder structure plays in DNA dynamics.

The question remains as to why linear crowders elongate DNA while branched crowders lead to compaction. Both conformations reduce the effective volume the DNA takes up in solution by either stretching into a long thin strand or compacting down to an ordered sphere. This crowding-induced phase transition arises from entropically-driven depletion interactions with the crowders. Namely, the DNA is forced to reduce its configurational volume to maximize the volume, and thus the entropy, of the crowders. However, why does the preferred configuration depend on crowder structure? As described in the Introduction, dextrans assume highly asymmetric random coil configurations in solution, compared to the hard sphere configurations of Ficoll and self-associated PEG crowders. If the DNA configurational transitions are entropically-driven then the DNA should assume the shape that allows for the most efficient packing of crowders around the DNA to maximize their available volume. Therefore, the DNA should assume a configuration that most closely matches that of the crowders while still reducing its volume from random coil. A previous simulation study comparing the effects of spherical vs cylindrical crowders on the conformational size of short rigid DNA molecules showed that cylindrical crowders elongated DNA while spherical crowders more readily compacted DNA [26]. While these results cannot be directly compared to our study, as our DNA is much larger and more flexible, they suggest that asymmetric crowders promote elongation to maximize the packing efficiency and available volume of the crowders. This study also showed that the presence of the DNA led to local nematic ordering of the crowders around the DNA which in turn facilitated DNA elongation [26]. In contrast, branched crowders are spherical in nature so packing is most efficient when DNA is likewise spherical.

## 3.2 Ionic Conditions

We next turn to the role of salt concentration on DNA crowding by carrying out measurements in the presence of NaCl concentrations that are *5x* higher and lower than our standard conditions (10 mM NaCl). Because our data show much more of a dependence on crowder structure than concentration, in these experiments we fixed our crowder concentration at 40% w/v. Figure 4A displays the measured diffusion coefficients for dilute and crowded conditions as a function of NaCl concentration. As shown, we see a surprising non-monotonic dependence of DNA mobility on [NaCl] when crowded: DNA exhibits the fastest diffusion at 10 mM NaCl. This effect is not seen in the dilute case in which the diffusion coefficient decreases with increasing NaCl concentration.

At first glance, the dilute case is counterintuitive in and of itself because as NaCl concentration increases, the wormlike chain model for DNA predicts that the persistence length $l_p$ should decrease [78]. This predicted dependence is due to $Na^+$ ions partially screening the negatively charged DNA backbone, reducing the repulsion between neighboring DNA segments and thereby making the DNA more flexible. According to the freely-jointed chain model for flexible polymers, increasing $l_p$, while keeping the overall polymer length $L$ fixed, should increase the radius of gyration of the polymer through the scaling, $R_G \sim (L*l_p)^{1/2}$ [79, 80]. Increasing $R_G$ (and thus $R_h$) should in turn reduce the diffusion coefficient in accordance with the Stokes-Einstein equation ($D \sim R_h^{-1}$). However, the Stokes-Einstein relation approximates polymers as hard spheres and ignores intramolecular fluctuations. As NaCl concentration increases, and the number of persistence lengths per chain likewise increases, the DNA becomes more flexible and has more conformational degrees of freedom. This increase in conformational degrees of freedom leads to increased intramolecular fluctuations, which can in turn lead to slower COM motion as thermal kicks are going more into internal DNA fluctuations and less into the motion of the molecule as a whole. This argument is supported by Figure 4B, which shows that the steady-state fluctuation range of DNA ($f_{ss}$) increases with increasing [NaCl] in dilute conditions. We also measured the spread in the major axis lengths ($\Delta R_{max}$) and find that, in the absence of crowding, $\Delta R_{max}$ increases with increasing [NaCl], indicating that higher salt conditions lead to more conformational states (Fig 4C). Finally, we find that $<R_{max}>$ is actually largest at high salt (50 mM) rather than low salt (2 mM) (Fig 4C), in contradiction with the classical scaling $R_G \sim (L*l_p)^{1/2}$, but in support of our measured dependence of $D$ on [NaCl], assuming the scaling $D \sim R_h^{-1}$.

Upon crowding, branched crowders induce greater DNA compaction and lead to smaller $f_{ss}$ values compared to linear crowders for all salt conditions (Fig 4B). Thus, the effects of crowder structure on DNA conformation (described in the previous section) appear robust to environmental conditions. However, we do find that salt concentration plays a strong role on DNA conformations induced by all crowders. High salt conditions (50 mM NaCl) enable all crowders to markedly suppress conformational fluctuations of DNA (seen as a large drop in $f_{ss}$ and $\Delta R_{max}$, Fig 4B,C) and compact DNA (seen as a large drop in $<R_{max}>$ and $<R_{min}>$, Fig 4D) to maximize their entropy. This large change is facilitated by the increased flexibility of DNA at high [NaCl]. Because DNA is more flexible it can more easily transition between different conformational states, so it can more readily undergo large entropically-driven changes to its conformation. This effect is in accord with the classic $\Psi$-compaction mechanism in which high salt conditions and crowding are both needed to efficiently induce bulk condensation of DNA [41]. In contrast, at the lowest NaCl concentration (2 mM) the steady-state fluctuation range remains nearly unchanged by crowders (Fig 4B). Low [NaCl] conditions also result in the smallest change to the conformational size of DNA, with DNA undergoing slight compaction in the presence of all crowder types (Fig 4D).

The question remains as to why the diffusion exhibits a non-monotonic dependence on [NaCl] when crowded. While 50 mM NaCl induces that largest change in DNA size from its dilute value, the most compacted and elongated DNA states are actually seen in 10 mM NaCl. As shown in Figure 4(C,D), the 10 mM NaCl data points lie at the extreme left and right-hand sides of the phase plots (low and high $<R_{max}>$ values) while the 2 mM and 50 mM NaCl data are mostly clustered together in the middle of the plot. As our previous data suggested [39, 40], these extreme conformational state changes are what enables DNA to move faster than expected among crowders.

## 4    Conclusion

We have investigated the role of crowder structure, size and concentration, as well as ionic conditions, on the diffusion and conformational dynamics of large DNA molecules. We crowded 115 kbp DNA with widely used synthetic crowders that can be categorized into linear (dextrans) and branched (Ficoll, PEG) structures of small (10 kDa) and large (~500 kDa) molecular weights. We present a number of intriguing results that have not been previously predicted or observed, and cannot be explained by classical polymer theory. In all crowding conditions, we find that DNA diffuses faster than classically expected based on the increasing viscosity of the solutions, which we attribute to changes to the random coil conformation DNA assumes in dilute conditions. We find that DNA elongates in the presence of linear crowders while branched crowders compact DNA. Elongated conformations undergo larger conformational fluctuations than dilute random coils, while compacted configurations are more spherical and access a smaller range of conformational states. These findings are largely independent of crowder concentration and molecular weight. We also find that DNA dynamics exhibit a complex interplay between salt concentration and crowding. In dilute conditions, DNA diffusion decreases and the range of conformational states increases with increasing salt (from 2 mM to 50 mM NaCl). However, upon crowding, DNA diffusion and conformational changes exhibit an emergent non-monotonic dependence on salt concentration, with DNA diffusing the fastest and exhibiting the most extreme compaction or elongation in 10 mM NaCl compared to lower and higher salt concentrations. Our collective results present several complex and unexpected phenomena that are highly relevant to polymer physics and cell biology alike. We hope that these results spur new theoretical investigations to fully understand these intriguing results.

## 5 Author Contributions

WMM prepared samples, collected data, analyzed data, interpreted data, wrote paper; SMG prepared samples, collected data, analyzed data; KR prepared samples, collected data; T-CW collected data, analyzed data; RMR-A designed experiments, analyzed data, interpreted data, wrote paper.

## 6 Funding

This research was funded by the Air Force Office of Scientific Research (Young Investigator Program Award No. FA95550-12-1-0315, Biomaterials Award No. FA9550-17-1-0249) and the National Institutes of Health (NNIGMS Award No. R15GM123420).

## 7 Conflict of Interest Statement

The authors declare no conflict of interest.

## 9    Figures and Captions

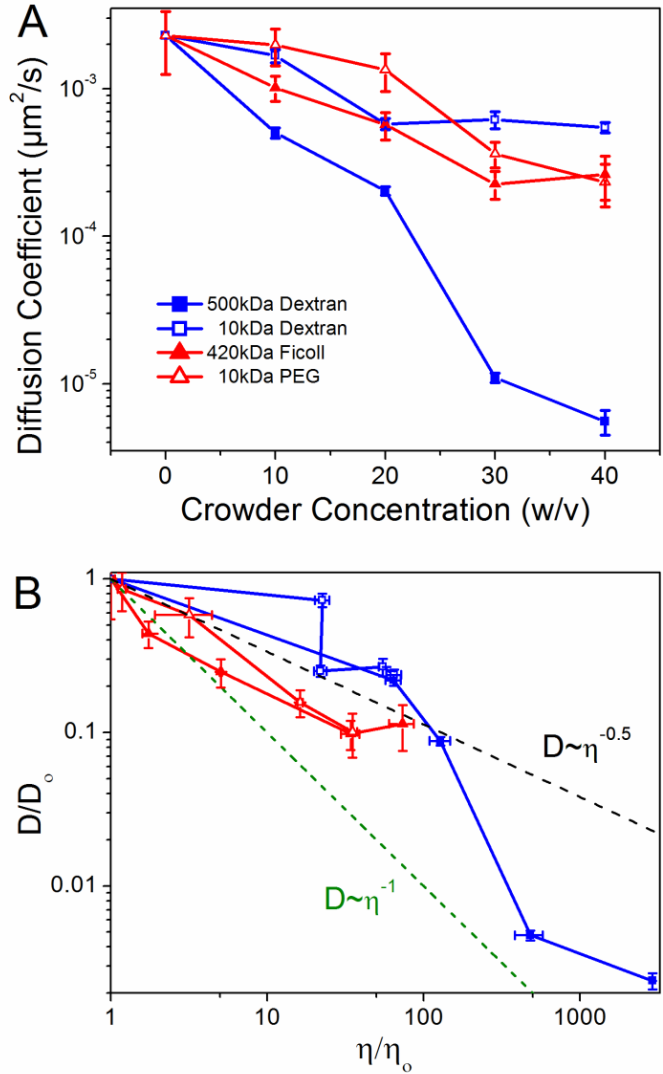

**Figure 1:** DNA diffusion decreases with increasing crowder concentrations, but universally diffuses faster than predicted by Stokes-Einstein scaling with viscosity. (A) Diffusion coefficients for DNA crowded by varying concentrations (w/v) of 500 kDa dextran (blue filled squares), 10 kDa dextran (blue open squares), 420 kDa Ficoll (red filled triangles) and 10 kDa PEG (red open triangles). (B) DNA diffusion coefficients, $D$, versus crowded solution viscosity, $\eta$, both normalized with respect to the corresponding dilute (0% crowding) values ($D_0$, $\eta_0$). The dashed green line indicates the Stokes-Einstein relationship $D\sim\eta^{-1}$, while the black dashed line shows the average empirical scaling, $D\sim\eta^{-0.5}$. Note that nearly every $D/D_o$ value lies above the predicted Stokes-Einstein value.

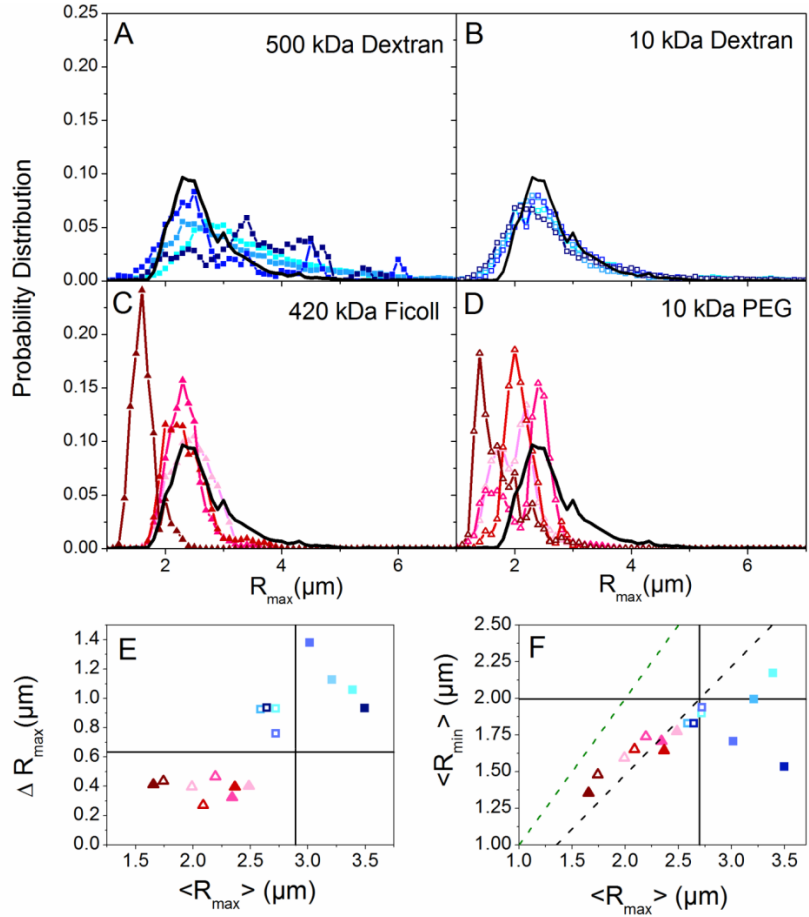

**Figure 2:** Crowding induces compaction or elongation of DNA dependent on the crowder structure. (A-D) Histograms of major axis lengths ($R_{max}$) for all crowding conditions compared to the dilute case (black line). Each panel displays results for a different crowder with blue hues denoting linear crowders (A,B; dextrans) and red hues indicating branched crowders (C,D; Ficoll, PEG). Closed symbols indicate high $M_w$ crowders (A,C; 500 kDa dextran, 420 kDa Ficoll) and open symbols denote low $M_w$ crowders (B,D; 10 kDa dextran, 10 kDa PEG). In each panel the color shade increases with increasing crowder concentration (10, 20, 30, 40% w/v). (E) The standard deviation of each $R_{max}$ distribution, $\Delta R_{max}$, versus the corresponding mean, $<R_{max}>$. The color scheme is the same as A-D. The dilute value is indicated by the intersection of the horizontal and vertical lines. Symbols above or below the horizontal line indicate increased or decreased ranges in conformational states accessed, and symbols to the left or right of the vertical line indicate crowding-induced compaction or elongation. (F) Phase plot of the mean major and minor axis lengths ($<R_{max}>$,$<R_{min}>$) for each crowded condition. The green dashed line denotes the $<R_{max}>$:$<R_{min}>$ ratio for a spherical particle. The black dashed line denotes $<R_{max}>$:$<R_{min}>$ for the empirically measured dilute random coil configuration. Note that branched crowders (triangles) generally decrease $<R_{max}>$ and $<R_{max}>$:$<R_{min}>$ indicating more spherical compacted conformations, while linear crowders (squares) tend to elongated DNA.

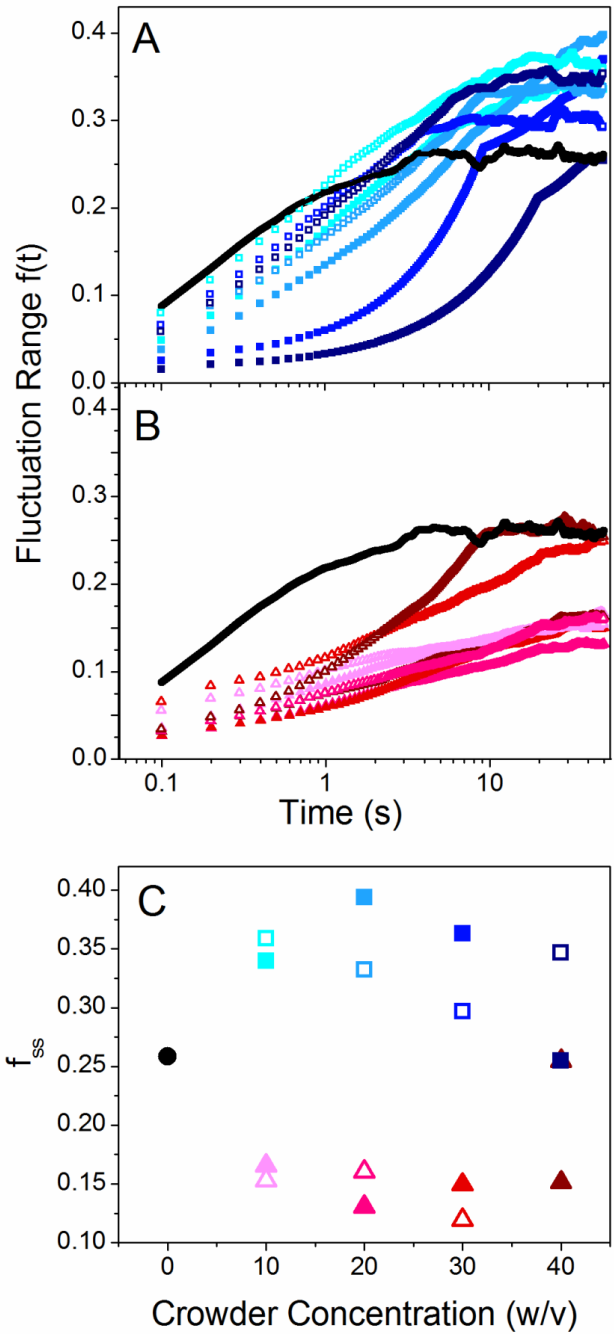

**Figure 3:** Crowded DNA molecules fluctuate over a range of different conformational states over time with a steady state fluctuation range that depends on crowder structure. (A,B) Fluctuation range, $f(t)$, as a function of time for linear (A) and branched (B) crowders compared to the dilute case (black lines). Open and closed symbols signify small (10 kDa) and large (~500 kDa) crowders and color shade increases with increasing crowder concentrations (10, 20, 30, 40% w/v). (C) Steady-state fluctuation range, $f_{ss}$ (i.e. the terminal plateau values of $f(t)$), versus crowder concentration for linear (squares) and branched (triangles) crowders. As shown, linear crowders increase $f_{ss}$ from the dilute value, while branched crowders reduce $f_{ss}$. These effects are largely independent of crowder size and concentration.

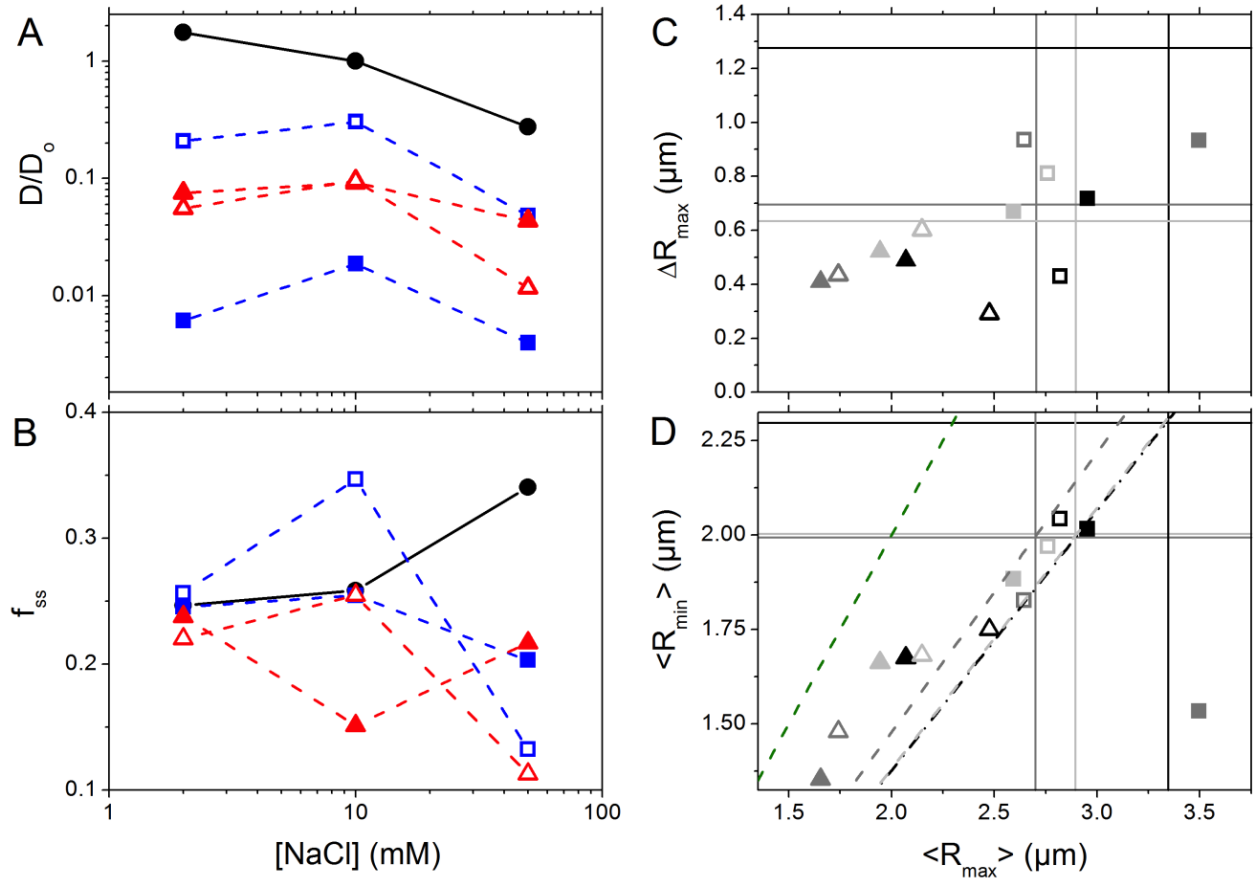

**Figure 4:** Diffusion and conformational dynamics of crowded DNA display unexpected non-monotonic dependence on NaCl concentration. (A) DNA diffusion coefficients, $D$, vs NaCl concentration for solutions with no crowders (black circles) and 40% w/v 500 kDa dextran (closed blue squares), 10 kDa dextran (open blue squares), 420 kDa Ficoll (closed red triangles) and 10 kDa PEG (open red triangles). Diffusion coefficients are normalized by the dilute $D$ value at 10 mM NaCl ($D_0$). (B) Steady-state conformational fluctuation range of DNA, $f_{ss}$, as a function of [NaCl]. Color scheme is as in A. (C) Phase plot of the mean major axis length $<R_{max}>$ and corresponding standard deviation $\Delta R_{max}$ for 10 kDa (open symbols) and ~500 kDa (closed symbols) linear (squares) and branched (triangles) crowders at NaCl concentrations of: 2 mM (light grey), 10 mM (grey), and 50 mM (black). The intersections of the color-coded horizontal and vertical lines denote the dilute values at the corresponding salt concentration. (D) Phase plot of the mean major and minor axis lengths ($<R_{max}>$, $<R_{min}>$) of crowded DNA. Colors and symbols are as in C. The dashed lines show the color-coded $<R_{max}>$:$<R_{min}>$ ratio (i.e. degree of sphericity) for the corresponding dilute DNA conformations. The green dashed line denotes the $<R_{max}>$:$<R_{min}>$ ratio for a spherical particle. As shown, 10 mM salt conditions lead to the most pronounced crowding-induced compaction and elongation (smallest and largest $R_{max}$) compared to lower or higher salt conditions.